\documentclass[useAMS,usenatbib]{mn2e}
\usepackage{times}
\usepackage{graphicx, latexsym, amssymb, amscd, psfrag}
\usepackage{epsfig}
\title[The mass assembly of galaxy groups]{The mass assembly of galaxy groups and the evolution of the magnitude gap}
\author[Dariush et al.]{Ali A. Dariush\thanks{E-mail: Ali.Dariush@astro.cf.ac.uk}$^{1,2}$,
Somak Raychaudhury$^{1}$, Trevor J. Ponman$^{1}$, Habib G. Khosroshahi$^{3}$,  
\newauthor 
Andrew J. Benson$^{4}$, Richard G. Bower$^{5}$, Frazer Pearce$^{6}$ \\
$^{1}$School of Physics and Astronomy, University of Birmingham, Birmingham B15~2TT, UK\\
$^{2}$School of Physics and Astronomy, Cardiff University, Queens Buildings, The Parade, Cardiff, CF24~3AA, UK\\
$^{3}$School of Astronomy, Institute for Research in Fundamental Sciences (IPM), P. O. Box 19395-5746, Tehran, Iran \\
$^{4}$California Institute of Technology, Pasadena, CA~91125, USA,\\
$^{5}$Department of Physics, University of Durham, South Road, Durham DH1~3LE, UK\\
$^{6}$School of Physics and Astronomy, University of Nottingham, Nottingham, NG7~2RD, UK}

\begin{document}

\pagerange{\pageref{firstpage}--\pageref{lastpage}} \pubyear{2010}

\maketitle

\label{firstpage}

\begin{abstract}

  We investigate the assembly of groups and clusters of galaxies using
  the Millennium dark matter simulation and the associated Millennium
  gas simulations, and semi-analytic catalogues of galaxies.  In
  particular, in order to find an observable quantity that could be
  used to identify early-formed groups, we study the development
  of the difference in magnitude between their brightest galaxies to
  assess the use of magnitude gaps as possible indicators.  We select
  galaxy groups and clusters at redshift $z\!=\! 1$ with dark matter
  halo mass $M(R_{\rm 200}) \geq 10^{13}\, h^{-1}\,$M$_{\odot}$, 
   and trace their properties until the present time ($z\!=\! 0$). 
   Further constraints are applied to keep those galaxy systems for which 
   the X-ray luminosity $L_{\rm X,bol} \ge 0.25\times 10^{42}\,h^{-2}$erg
  s$^{-1}$ At redshift $z\!=\! 0$.  
  While it is true that a large magnitude gap between the two
  brightest galaxies of a particular group often indicates that a
  large fraction of its mass was assembled at an early epoch, it is
  not a necessary condition.  More than 90\% of fossil groups defined
  on the basis of their magnitude gaps (at any epoch between $0<z<1$ )
  cease to be fossils within 4~Gyr, mostly because other massive
  galaxies are assembled within their cores, even though most of the
  mass in their haloes might have been assembled at early times.  We
  show that compared to the conventional definition of fossil galaxy
  groups based on the magnitude gap $\Delta m_{\rm 12}\!\geq\! 2$ (in
  the $R$-band, within $0.5 \, R_{\rm 200}$ of the centre of the
  group), an alternative criterion $\Delta m_{\rm 14} \geq 2.5$
  (within the same radius) finds 50\% more early-formed systems, and
  those that on average retain their fossil phase longer.
   However, the conventional criterion performs marginally
  better at finding early-formed groups at the high-mass end of
  groups. Nevertheless, both criteria fail to identify a majority
  of the early-formed systems.

\end{abstract}


\section{Introduction}

Although existing observations of the large scale structure of the
Universe overwhelmingly favour cold dark matter cosmologies with
hierarchical structure formation, the paradigm faces challenges both
from the existence of luminous passive galaxies at high redshift, and
the abundance of low-mass galaxies in the local universe
\citep[e.g][]{baugh06,Balogh08}. Galaxies dominate the visible
universe and any cosmological model is expected to reproduce the
observed global properties of galaxies, at least statistically, in the
first instance. A significant fraction of the evolutionary life of
many galaxies is spent in the environment of small systems
(i.e. groups), where close interactions and mergers of galaxies occur
with higher efficiency than in massive haloes such as galaxy clusters
\citep[e.g.][]{miles04}.  The observable properties of the baryonic
content of a group, which consists of the constituent galaxies and
the inter-galactic medium (IGM), should be linked to mass assembly
of the host group and its subsequent evolution.  Among such observable
properties, the ''magnitude gap'', i.e., the difference in the
magnitudes of the two brightest galaxies, has been widely used as an
optical parameter related to mass assembly of groups and clusters.

Several studies show that a system of galaxies, where most of the mass
has been assembled very early, develops a larger magnitude gap
compared to systems that form later.  The idea is supported in
observational samples \citep[e.g.][]{Ponman94,habib04,Habib07},
theoretical studies \citep[e.g.][]{Milos06,Van07} and in detailed
analysis of N-body numerical simulations
\citep{Barnes89,donghia05,Dariush07}.  These studies also predict that
such early-formed galaxy groups or clusters should be relaxed and
relatively more isolated systems in comparison to their later-formed
counterparts.  

\citet{Jones03} defined such early-formed systems (also
known as {\it fossils}) to have a minimum X-ray luminosity of $L_{\rm
  X,bol} \geq 0.25 \times 10^{42} h^{-2}$erg s$^{-1}$, and a large
magnitude gap in the $R$-band between their first two brightest
galaxies, i.e. $\Delta m_{\rm 12} \geq 2.0$, to distinguish them from
late-formed groups and clusters.  Recent research based on the Sloan
Digital Sky Survey has either used only the optical criterion
\citep{Santos07} or both optical and X-ray criteria
\citep{Eigen09,Voevo09,labarb09} to identify fossils. In the former
case, if the optical definition is solely employed, the chance of
identifying truly early-formed systems diminishes, since a large
fraction of the systems detected might be in the stage of collapsing
for the first time, and so would not be X-ray luminous.

From numerical simulations, \citet{vbb08} found that $\Delta m_{\rm
  12}$ may not be a good indicator for identifying early-formed
groups, since the condition would no longer be fulfilled when a galaxy
of intermediate magnitude fell into the group. This study was based on
simulations of dark matter particles only, and does reveal how
frequently such a situation would arise. Other, potentially more
robust, magnitude gap criteria have been considered in the literature.
For example, \citet{Sales07} finds that the difference in magnitude
between the first and 10$^{th}$ brightest galaxies in three fossil
groups span a range of $\sim$3--5, in agreement with their results
from the analysis of the Millennium data together with the
semi-analytic catalogue of \citet{Croton06}.

The overall number of observed fossil galaxy groups is small, making
it difficult for observed systems to be statistically compared to
simulated systems.  In spite of their low space density, fossils have
been used to test models of cosmological evolution
\citep{Milos06,Habib07,vbb08,dm08}, since the criteria for
observationally identifying such systems are simple
\citep{Jones03,habib06}, and it is generally assumed that they are the
archetypal relaxed systems, consisting of a group-scale X-ray halo,
the optical image being dominated by a giant elliptical galaxy at the
core \citep{Ponman94}. Indeed, if they are relaxed early-formed
systems, fossil groups can be the ideal systems in which to study
mechanisms of feedback, and the interaction of central AGN and the IGM
of the group, since the effect of the AGN would not be complicated by
the effect of recent mergers \citep{jetha08,jetha09}. The lack of
recent merging activity would also predict the absence of current or
recent star formation in early-type galaxies belonging to fossil
groups \citep[e.g.][]{nolan07}, and the relative dearth of red
star-forming galaxies compared to similar elliptical-dominated
non-fossil groups and clusters \citep{sm09}.

The use of cosmological simulations in the study of the evolution of
galaxy groups will have to employ a semi-analytic scheme for simulating
galaxies, and the results will be dependent on the appropriate
characterisation of the models that describe galaxy formation and
evolution.  Once the hierarchical buildup of dark matter haloes is
computed from N-body simulations, galaxy formation is modelled by
considering the rate at which gas can cool within these haloes. This
involves assumptions for the rate of galaxy merging (driven by
dynamical friction) and the rate and efficiency of star formation and
the associated feedback in individual galaxies
\citep{Croton06,Bower06}. 

In a previous paper \citep{Dariush07}, we studied the formation of
fossil groups in the Millennium Simulations, and showed that the
conventional definition of fossils (namely a large magnitude gap
between the two brightest galaxies within half a virial radius and a
lower limit to the X-ray luminosity $L_{\rm X,bol} \geq 2.5 \times 10^{42}
h^{-2}$erg s$^{-1}$), results in the identification of haloes which
are $\approx ~$10\%-20\% more massive than the rest of the population
of galaxy groups with the same halo mass and X-ray luminosity, when
the Universe was half of its current age. This clearly indicates an
early formation epoch for fossils. In addition, it was shown that the
conventional fossil selection criteria filter out spurious systems,
and therefore there is a very small probability for a large magnitude
gap in a halo to occur at random. The fraction of late-formed systems
that are spuriously identified as fossils was found to be $\approx
~$4--8\%, almost independent of halo mass \citep{Dariush07,Smith09}.
Another important outcome of the \citet{Dariush07}
study was the consistency between
the space density of fossils found in the simulations and that from
observational samples.

Although the results from this previous analysis of the Millennium
simulations are shown to be in fair agreement with observation, we did
not investigate the evolution of the magnitude gap in either fossil or
control groups with redshift. Furthermore, the number of (fossil or
control) groups used was small ($\sim$400), which did not allow us to
fully explore the connection between the halo mass and magnitude gap
in such systems.

In this paper, we select early-formed galaxy groups from the
Millennium simulations, purely on the basis of their halo mass
evolution from present time up to redshift $z \approx 1.0$, and, with
the help of associated semi-analytic catalogues, study the evolution
of the magnitude gap between their brightest galaxies. Our aim is to
(a) investigate how well the conventional optical
selection criterion, namely the $\Delta m\ge 2$ gap between the two
brightest galaxies, is able to identify early-formed galaxy groups, and (b)
to find whether we can find a better criterion to identify groups
that have assembled most of their mass at an early epoch.

In \S2, we describe the various simulation suites used in this work,
and in \S3 the data we extract from them. In \S4, we study in detail
the evolution with epoch of various measurable parameters for a large
sample of early-formed ``fossil systems'' and two comparable sample of
control systems, and compare these properties. In \S5, we examine the
case for a revision of the criteria to observationally find fossil in
order to ensure a higher incidence of genuine early-formed systems. We
summarise our conclusions in \S6. We adopt $H_0 = 100\, h$
km~s$^{−1}$ Mpc$^{−1}$ for the Hubble constant, with $h=0.73$.


\section{Description of the Simulations}

\subsection{The Millennium Simulation}

The Millennium Run consists of a simulation, in a Universe consistent
with concordance $\Lambda$CDM cosmology, of 2160$^3$ particles of
individual mass $8.6\times10^{8}h^{-1}$ M$_{\odot}$, within a
co-moving periodic box of side 500$h^{-1}$ Mpc, employing a
gravitational softening of 5$h^{-1}$ kpc, from redshift $z=127$ to the
present day \citep{Springel05}.  The basic setup is that of an
inflationary Universe, dominated by dark matter particles, leading to
a bottom-up hierarchy of structure formation, which involves the
collapse and merger of small dense haloes at high redshifts, into the
modern-day observed large virialised systems such as groups and
clusters.  The cosmological parameters used by the Millennium Simulation
were $\Omega_\Lambda = 0.75$, $\Omega_M = 0.25$, $\Omega_b = 0.045$, 
$n = 1$, and $\sigma_8 = 0.9$, and the Hubble parameter 
$h = 0.73$.

Dark matter haloes are found in this simulation down to a resolution
limit of 20 particles, yielding a minimum halo mass of 1.72$\times
10^{10}h^{-1}$ M$_{\odot}$. Haloes in the simulation are found using a
friends-of-friends (FOF) group finder, configured to extract haloes
with overdensities of at least 200 relative to the critical density
\citep{Springel05}.
Within a FOF halo, substructures or subhaloes are identified using the
SUBFIND algorithm developed by \citet{Springel01}, and the treatment
of the orbital decay of satellites is described in the next section.

During the Millennium Simulation, 64 time-slices of the locations and
velocities of all the particles were stored, spread approximately
logarithmically in time between $z=127$ and $z=0$. From these
time-slices, merger trees were built by combining the tables of all
haloes found at any given output epoch, thus enabling us to
trace the growth of haloes and their subhaloes through time within the
simulation.

\subsection{Semi-analytic galaxy catalogues} 

\subsubsection{The Croton et al. semi-analytic catalogue}
\label{SAMcroton}

\citet{Croton06} simulated the growth of galaxies, and their central
supermassive black holes, by self-consistently implementing
semi-analytic models of galaxies on the dark matter haloes of the
\citet{Springel05} simulation, Their semi-analytic catalogue contains
9 million galaxies at $z=0$ brighter than absolute magnitude $M_R\!-\!5
\log\, h = -16.6$, ``observed'' in $B$, $V$, $R$, $I$ and $K$
filters. The models focus on the growth of black holes and AGN as
sources of feedback sources.  The inclusion of AGN feedback in the
semi-analytic model (allowing central cooling to be suppressed in
massive haloes that undergo quasi-static cooling), and its good
agreement with the observed galaxy luminosity function, distribution
of galaxy colours and of the clustering properties of galaxies, make
this catalogue suitable for our study.

In this semi-analytic formulation, galaxies initially form within
small dark matter haloes.  Such a halo may fall into a larger halo as
the simulation evolves.  The ``galaxy'' within this halo then becomes
a satellite galaxy within the main halo, and follows the track of its
original dark matter halo (now a subhalo), until the mass of the
subhalo drops below 1.72$\times 10^{10}h^{-1}$ M$_{\odot}$. This limit
corresponds to the 20-particle limit for dark haloes in the original
Millennium Simulation. At this point the galaxy is assumed to spiral
into the centre of the halo, on some fraction of the dynamical
friction timescale, where it merges with the central galaxy of the
larger halo \citep{Croton06}.

\subsubsection{The Bower et al. semi-analytic catalogue}
\label{SAMbower}

The \citet{Bower06} model also makes use of the Millennium Simulation,
but utilises merger trees constructed with the algorithm described by
\citet{Harker06}. The cooling of gas and the subsequent formation of
galaxies and black holes is followed through the merging hierarchy of
each tree utilising the {\sc Galform} semi-analytic model
\citep{Cole00,Bower06,Malbon07}. At $z=0$ this results in 4,491,139
galaxies brighter than a limiting absolute magnitude of $M_K\!-\!5
\log\, h =-19.4$.

In addition to feedback from supernovae, the
\citet{Bower06} model accounts for energy input from AGN, resulting in
a suppression of cooling in the hot atmospheres of massive haloes. The
resulting galaxy population is in excellent agreement with the
observed $z=0$ galaxy luminosity function in $B$ and $K$ bands, the
$z=0$ colour distribution and also with the evolution of the galaxy
stellar mass function from $z=0$ to $z\approx 5$. This model is
therefore similarly well-suited to our study of fossil systems. If a
halo in a merger tree has multiple progenitors, all but the most
massive are considered to become subhaloes orbiting within the larger
host halo and any galaxies they contain therefore become satellite
galaxies in that halo.

Due to the limited resolution of the Millennium Simulation (which may
cause dynamical friction timescales to be poorly estimated), the time
between becoming a satellite and merging with the central galaxy of
the halo is computed from the analytic dynamical friction
timescale. Specifically, each new satellite is randomly assigned
orbital parameters from a distribution measured from N-body
simulations and the appropriate dynamical friction timescale computed
following the approach of \citet{Cole00}, but multiplied by a factor
of 1.5. This was found to produce the best fit to the luminosity
function in \protect\citet{Bower06} but is also in good agreement with
the results of \protect\citet{Boylan07} who compared the predictions
of analytic dynamical friction timescales with those from idealised
N-body simulations.  The satellite is allowed to orbit for the period
of time calculated above, after which it is merged with the central
galaxy of the halo. If the host halo doubles in mass before a
satellite can merge, the satellite orbit is assumed to be reset by the
merging which lead to that mass growth and so a new set of orbital
parameters are assigned and a new merging timescale computed.

\subsection{The Millennium Gas Simulations}

The Millennium Gas Simulations are a series of hydrodynamical models
constructed within the same volume, and values of initial perturbation
amplitudes and phases, as the parent dark-matter-only Millennium
Simulation \citep[see, e.g.,][]{hartley08}. Of the three principal
models completed in this work, each contains additional baryonic
physics: (i) the first does not follow the effects of radiative
cooling and so overpredicts the luminosities of group-scale objects
significantly, (ii) the second includes a simple preheating scheme
that is tuned to match the observed X-ray properties of clusters at
the present day and (iii) the third includes a simple feedback model
that matches the observed properties of clusters today. We have used
the second of these models in this work, as we only utilise the
hydrodynamical properties of the groups at $z=0$, where the
observational and simulation results are well matched.

The Millennium Gas Simulations consist of $5 \times 10^8$ particles of
each species, resulting in a dark matter mass of $1.422 \times
10^{10}h^{-1}$ M$_\odot$ per particle and a gas mass of $3.12 \times
10^{9}h^{-1}$ M$_\odot$ per particle. The Millennium Simulation has
roughly 20 times better mass resolution than this and so some
perturbation of the dark matter halo locations is to be expected. In
practice the position and mass of dark matter haloes above
$10^{13}h^{-1}$ M$_\odot$ are recovered to within $50\,h^{-1}$kpc
between the two volumes, allowing straightforward halo-halo matching
in the large majority of cases.

The Millennium gas simulations used exactly the same cosmological
parameters as those of the dark matter simulations.  With the
inclusion of a gaseous component, additional care needs to be taken in
choosing the gravitational softening length in order to avoid spurious
heating \citep{Steinmetz97}. We use a comoving value of
$25(1+z)h^{-1}$ kpc, roughly 4\% of the mean inter-particle separation
\citep{Borgani} until $z=3$, above which a maximum comoving value of
$100h^{-1}$ kpc is adopted. A different output strategy is followed in
the Millennium Gas Simulations, where the results are output uniformly
in time with an interval roughly corresponding to the dynamical time
of objects of interest. This strategy results in 160 rather than 64
outputs and places particular emphasis on the late stages of the
simulation.

\section{Datasets used in this work }
\label{Data}

We start with a catalogue of groups extracted by the friends-of-friends
(FoF) algorithm employed in the Millennium Dark matter runs. Hereafter a
``group'' or ``group halo'' would refer to a group taken from this
catalogue.  In order to follow the evolution of groups from $z\!\sim$1
to the present epoch, we have to combine various sets of information
from this FoF group catalogue and the associated semi-analytic
catalogues of galaxies, as well as the gas simulations.

We select all groups of $M(R_{\rm 200}) \!\geq\!  10^{13}\,
h^{-1}\,$M$_{\odot}$ from the FoF group catalogue at $z=0.998$. The
mass cut-off is intended to ensure that the progenitors of the present
day galaxy groups are indeed groups at $z \sim$1 with at least four or
five members (galaxies), above the magnitude cut of the catalogue.
The evolution of each group was followed from $z=0.998$ to $z=0$ (at
23 discrete values, equally spaced in $\log z$) by matching the
position of each halo to its descendants at later redshifts.

The position of the central galaxy of each galaxy group, and the
corresponding dark matter halo, were used to identify the member
galaxies of each group. At each redshift and for each group halo,
optical properties were extracted for its corresponding galaxies from
the semi-analytic galaxy catalogue. The model galaxies become
incomplete below a magnitude limit of M$_{K}-5\log(h) \sim -19.7$, due
to the limited mass resolution of the Millennium simulation. We
applied a $K$-band absolute magnitude cut-off of M$_K \lesssim -19$ on
galaxies at all redshifts.  During the matching process, for more than
99\% of the groups at each redshift, corresponding galaxies were found
in the semi-analytic galaxy catalogue. The remaining groups were
excluded from our final compiled list.

In order to find the gas properties of all groups at $z\! =\! 0$, we
cross-correlated our list of groups with the Millennium gas
catalogue, and find the bolometric X-ray luminosity of our selected
groups at $z$=0.  

\begin{figure}
\epsfig{file=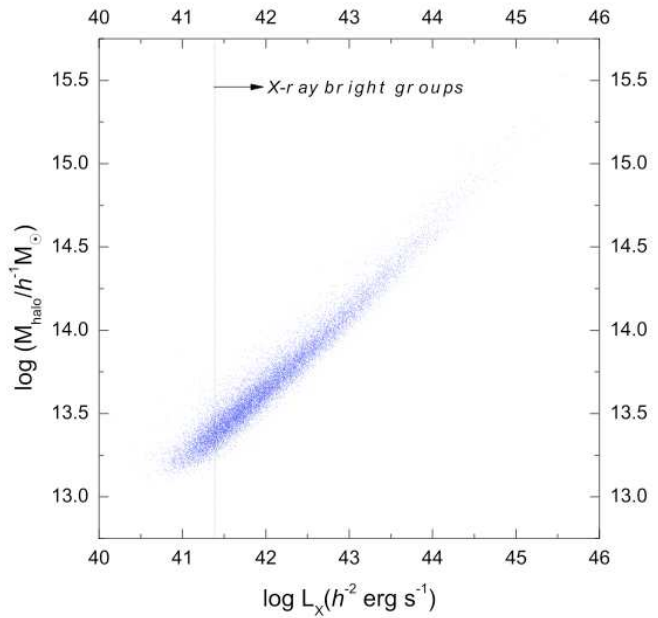,width=1.05\hsize}
\caption{The relation between the mass of group haloes (within $R_{\rm
    200}$) at $z=0$ from the Millennium DM simulation, and the
  bolometric X-ray luminosity of the corresponding haloes in the
  Millennium gas simulation. All groups have $M(R_{\rm 200}) \geq
  10^{13}\, h^{-1}\,$M$_{\odot}$ at $z\sim1.0$.  The {\it vertical
    dashed-line} corresponds to the X-ray luminosity threshold $L_{\rm
    X,bol} = 0.25\times 10^{42}\,h^{-2}$erg s$^{-1}$ generally adopted
  to define fossil groups (see Sec.~\ref{fossil}). Of the 17866 groups
  matched in the two catalogues, 14628 groups lie above this
  threshold. In this paper, we call these ``X-ray bright groups''.  }
\label{alifig01}
\end{figure}

Out of 19066 dark matter group haloes with $M(R_{\rm 200}) \geq
10^{13}\, h^{-1}\,$M$_{\odot}$ selected at $z=0.998$, optical
properties from the semi-analytic catalogue (as well as gas properties
from the gas simulations at $z$=0) and the entire history of evolution
at all redshifts up to $z$=0, were found for 17866 ($\sim$94\% of the
initial sample at $z\sim1$) of group haloes. Fig.~\ref{alifig01} shows
the bolometric X-ray luminosity from the Millennium gas simulation,
plotted against the corresponding dark matter halo mass of each
group, at redshift $z$=0 for all of the matched 17866 groups. 

The vertical dashed line in Fig.~\ref{alifig01} corresponds to the
conventional X-ray luminosity threshold ($L_{\rm X,bol} = 0.25\times
10^{42}\,h^{-2}$erg s$^{-1}$) for fossil groups \citep{Jones03}, as
adopted in Sec.~\ref{fossil} to define X-ray bright groups. There are
14628 groups above this threshold, out of 17866 groups.  These X-ray
bright groups will constitute the main data set for the rest of our
analysis, except for \S\ref{SAM} and \S\ref{Delta4}, where the whole
range of halo mass will be explored to study the magnitude gap
statistics and the local environment of groups,

\section{Results}
\label{results}

\subsection{The $R$-band Magnitude Gap Statistic}
\label{SAM}

The dynamical friction $f_{\rm dyn}$
will cause the more luminous galaxies in a group to merge on a time scale
which depends upon the velocity dispersion of the group, and 
since $f_{\rm dyn} \propto v^{-2}$ this 
is more frequent in poorer groups than in clusters
\citep[e.g.][]{miles04,miles06}. As a result, on group scales, the 
likelihood of a few of the brightest galaxies merging to form the
brightest galaxy, leading to a large magnitude gap within a Hubble
time, is higher.
Thus, the distribution of the magnitude gap between the brightest galaxy,
and the second and third brightest galaxies, in each group, 
is often used as an indicator
the dynamical age
of group, particularly in fossil groups
\citep{Milos06,Van07,Dariush07,vbb08}.

We determine the magnitude gaps from the Millennium semi-analytic
models of \citet{Bower06} and \citet{Croton06}, and compare them with
observational results from the Sloan Digital Sky Survey (SDSS) C4
cluster catalogue data of \citet{Miller05} and the 2-degree Field
Galaxy Redshift Survey (2dFGRS) group catalogue of \citet{Van07}.

The 2dFGRS group catalogue is constructed based on a halo-based group
finder algorithm of \citet{Yang05} and contains $\sim 6300$ groups
within the mass range $\log (M/h^{-1}\,$M$_{\odot}) \geq 13.0$ where
the mass of each group has been determined from the total luminosity
of all group members brighter than $M_{b_j} -5\log h=-18$.
The C4 catalogue \citet{Miller05} consists of $\sim 730$ clusters
identified in the spectroscopic sample of the Second Data Release
(DR2) of the SDSS inside the mass range $13.69 \leq \log
(M/h^{-1}\,$M$_{\odot}) \leq 15.0$, estimated from the total $r$-band
optical luminosity of cluster galaxies. 

\begin{figure*}
\begin{center}
\epsfig{file=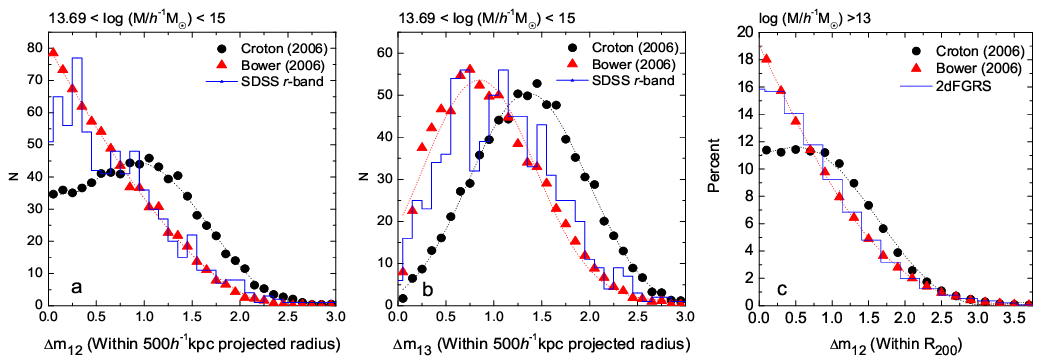,width=1.07\hsize}
\caption{The $R$-band magnitude gap distribution for haloes from the
Millennium semi-analytic models of \citet{Bower06} ({\it red
triangles}) and \citet{Croton06} ({\it black circles}) superposed on
the data from 2dFGRS group catalogue of \citet{Van07} as well as SDSS C4 cluster
catalogue of \citet{Miller05} ({\it blue histograms}).  
(a) The
magnitude gap $\Delta m_{\rm 12}$ 
between the
the first and second most luminous
galaxies, compared with galaxies from the SDSS C4
catalogue of clusters computed within projected radius of
500$h^{-1}$kpc.  (b) The same as in (a) but for 
the 
magnitude gap $\Delta m_{13}$
between the first and the third most luminous
galaxies.
(c) The magnitude gap $\Delta m_{\rm 12}$ estimated within $R_{\rm 200}$,
 compared with galaxies from the
2dFGRS group catalogue.
The $\sim 6300$ 2dFGRS groups are within the mass range
$\log (M(R_{\rm 200})/h^{-1}\,$M$_{\odot}) \geq 13.0$,
and those from SDSS
C4 catalogue consist of $\sim 730$ clusters within mass range $13.69
\leq \log (M(R_{\rm 200})/h^{-1}\,$M$_{\odot}) \leq 15.0$.  
}
\label{alifig02}
\end{center}
\end{figure*}

The results of the comparison between our distribution of the
estimated $R$-band magnitude gaps from semi-analytic models of
\citet{Bower06} as well as \citet{Croton06}, based on the Millennium
simulation (red triangles and black circles respectively), and the
observed results from the C4 cluster catalogue and the 2dFGRS group
catalogues (blue histogram), are shown in Fig.~\ref{alifig02}.  The
magnitude gap statistics $\Delta m_{\rm 12}$ and $\Delta m_{13}$ from
\citet{Bower06} are in excellent agreement with those obtained from
2dFGRS group catalogue and SDSS C4 catalogue of clusters. However, the
semi-analytic galaxy catalogue of \citet{Croton06} predicts a larger
fraction of groups with $\Delta m_{\rm 12} \geq 2.0$ for both the SDSS
and 2dFGRS samples. This is in particular of great importance to the
determination of the space density of fossil galaxy groups,
and the comparison of fossil
samples drawn from simulated and observed catalogues,
which use
the magnitude gap as a key discriminant, 
\citep[e.g., see
Table~1 of][]{Dariush07}.

The shift of the distribution of $\Delta m_{\rm 12}$ and $\Delta
m_{13}$ to larger values than observed in the \citet{Croton06} model
may reflect the fact that in the \citet{Bower06} model, the treatment
of dynamical friction differs from that used by \citet{Croton06}. In
both models, N-body dynamics are used to follow the orbital decay of
satellite galaxies whose subhalo can be resolved.  However, when the
subhalo can no longer be reliably followed, the dynamical friction
calculations differ. In the \citet{Bower06} model, the dynamical
friction timescale is initially calculated following
\citet{Cole00}. However, if the host halo of the satellite is deemed
to undergo a formation event (corresponding to a mass doubling since
the previous formation event), before the satellite merges, then a new
orbit for the satellite is selected at random, and the dynamical
friction timescale for the satellite is recalculated.  This
calculation takes into account the scattering of galaxies to larger
energy orbits during the merger of their parent halo.

Another possible cause for the success of the \citet{Bower06} model in
matching the magnitude gap statistics is that it predicts a large
scatter in the relation between galaxy stellar mass and halo
mass---significantly more than in the \citet{Croton06} model.  As a
result, sometimes rather large satellite haloes arrive carrying
relatively small galaxies resulting in a large difference between the
magnitude of the dominant object and the next most luminous.  This
difference occurs because the AGN feedback is not guaranteed to switch
off the cooling at a particular halo mass in the \citet{Bower06}
model, as it depends on the merging and cooling history of each halo.

For the purposes of this work, the fact that the \citet{Bower06}
method for computing merging timescales results in good agreement with
the observed magnitude gap distributions, makes it well suited for the
remainder of our study.

\subsection{Evolution of galaxy groups}

In cosmological simulations, the age of galaxy groups can be expressed
in terms of the rate of the mass assembly of the groups. This means
that for a given group halo mass, groups that formed early, assemble
most of their masses at an earlier epoch in comparison to younger
groups. Thus the {\it assembly time} of a dark matter halo, defined as
the look-back time at which its main progenitor reaches a mass that is
half of the halo's present day mass, is larger in ``older'' systems
than in ``younger'' ones. Of course, in cosmological simulations such
as the Millennium runs, where the structures in Universe from
hierarchically, massive systems which form later turn out to have
shorter assembly time than low mass groups. Therefore one should take
into account the mass of systems when comparing the mass assembly of
various types of groups and clusters.
  
\subsubsection{Fossil groups of galaxies}
\label{fossil} 

How is the history of mass assembly of a group or cluster related to
its present observable parameters? It is expected that groups which
have formed earlier tend to be more dynamically relaxed, thus
resulting in a hotter intergalactic medium (IGM) and being more likely
to be X-ray luminous \citep{miles04,forbes06}. It has been shown that
in X-ray luminous systems, the brightest galaxies are bigger and more
optically luminous and those belonging to systems that have little or
no diffuse X-ray emission \citep[e.g.,][]{habib-gems}.  On the other
hand, groups with the same mass that have formed late, and are still
in a state of collapse, would not show X-ray emission associated with
their IGM \citep{Jesper06,bai10}, and are less likely to be dominated by a
massive elliptical in their cores.

Hitherto the so-called {\it fossil galaxy groups}, which are supposed
to be canonical examples of groups that have formed early, have been
identified by requiring that their X-ray luminosities exceed $L_{\rm
  X,bol} \geq 0.25 \times 10^{42} h^{-2}$erg s$^{-1}$
\citep[e.g.][]{Habib07,Jones03}. In addition, a fossil group needs to
have, within half a virial radius of the group's centre, the second
brightest galaxy to be at least 2-mag fainter than the brightest
galaxy, i.e. $\Delta m_{\rm 12} \geq 2.0$ \footnote{This condition can
  be replaced by $\log (L_2/L_1) \leq -0.8$ where $L_1$ and $L_2$ are
  the luminosities of the first two brightest galaxies.}.  So far
these two observational criteria have been jointly used to explore
fossil groups and clusters of galaxies. Therefore,
Fig.~\ref{alifig03}, which displays all the {\it X-ray bright groups}
fulfilling the X-ray criterion in Fig.~\ref{alifig01}, and the optical
criterion $\Delta m_{\rm 12} \geq 2.0$ (dotted horizontal line) should
separate groups which have been formed earlier in comparison to their
counterparts with $\Delta m_{\rm 12}<2.0$.

Note that, in numerical simulations, fossils are identified as groups
with $\Delta m_{\rm 12} \geq 2.0$ within $R_{\rm 200}$ or $0.5R_{\rm
  200}$. Our results from this study as well as those represented in
\citet{Dariush07} show that the fraction of fossils (and therefore
their space densities) depend on the search radius within which
$\Delta m_{\rm 12}$ is estimated, whereas the history or mass assembly
does not change that much.

\subsubsection{The mass assembly of X-ray fossil groups}
\label{evolution1}

Let us introduce the parameter $\alpha_z$ which for an individual group is
the ratio of its mass at redshift $z$ to its final mass at $z=0$,
i.e. $\alpha_z\equiv M_{z}/M_{z=0}$. 
Thus at a given redshift $z$, groups with larger
$\alpha_z$ have assembled a larger fraction of their final mass by
$z$ than groups with smaller values of $\alpha_z$.

\begin{figure*}
\epsfig{file=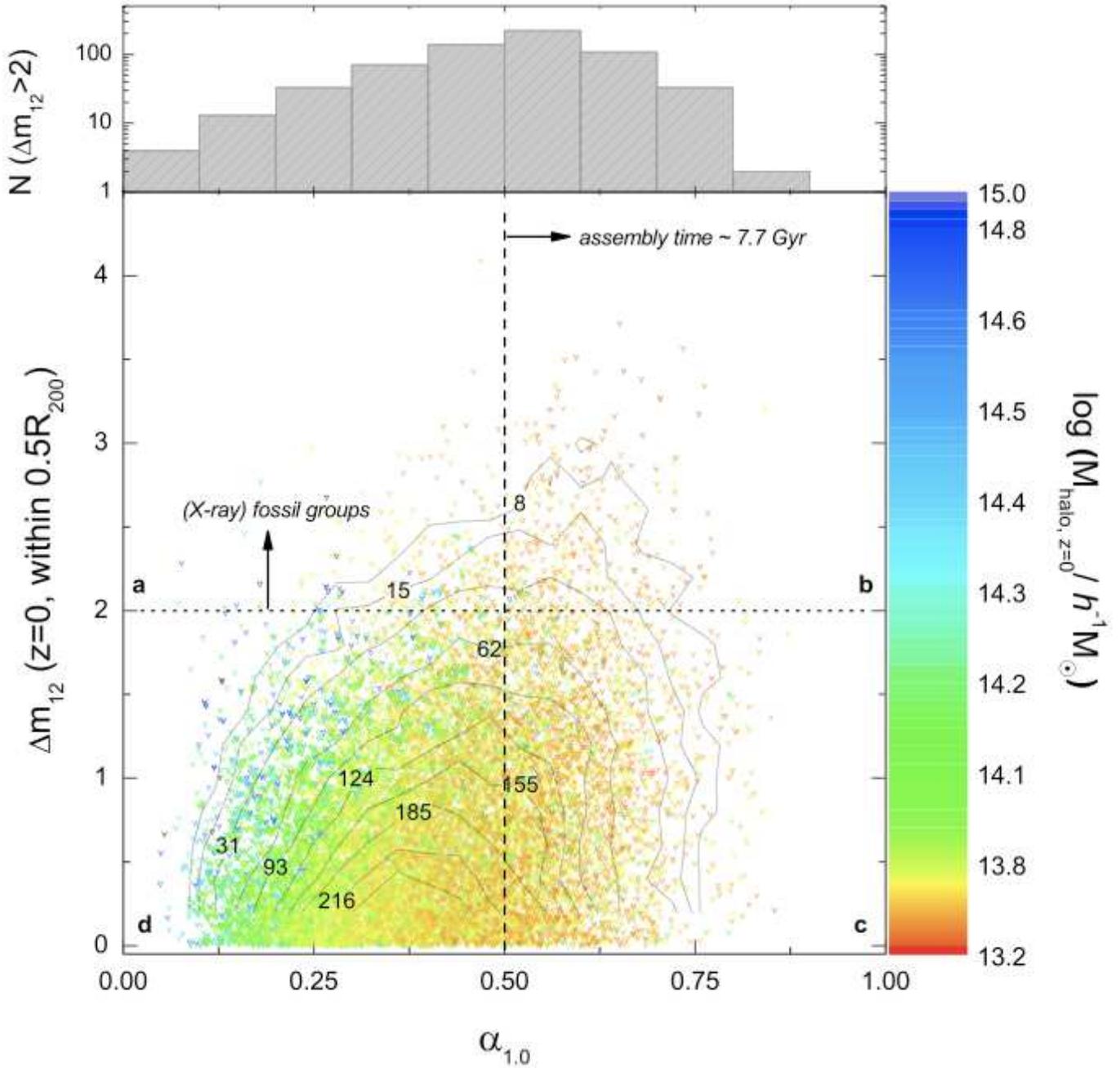,width=1.05\hsize}
\caption{ The magnitude gap $\Delta m_{\rm 12}$ within $0.5\,R_{\rm
    200}$, estimated for all 14628 X-ray bright groups in
  Fig.~\ref{alifig01} (i.e. groups with $L_{\rm X,bol} \geq 0.25
  \times 10^{42} h^{-2}$erg s$^{-1}$) 
at $z=0$ versus the ratio of the
  group halo mass at redshift $z=1$ to its mass at $z=0$
  ($\alpha_{1.0}$). The {\it horizontal dashed-line} separates groups
  into fossils ($\Delta m_{\rm 12} \geq 2.0$) and non-fossils ($\Delta
  m_{\rm 12} < 2.0$). The {\it vertical dashed-line} corresponds to
  $\alpha_{1.0}=0.5$. Groups with $\alpha_{1.0} \geq 0.5$ have formed
  more than half of their mass by $z \sim 1.0$ and hence have a
  minimum assembly time of about $\sim 7.7$ Gyr. Data points are
  colour-coded according to FoF group halo mass $M_{R200}$ at present
  epoch. The density of data points is represented by {\it black
    contour lines} which is the number of groups in each of $25 \times
  25$ cells of an overlaid grid, equally spaced horizontally and
  vertically. The {\it upper panel} represents the histogram of X-ray
  bright fossil groups, i.e. all groups with $\Delta m_{\rm 12} \geq
  2.0$ and $L_{\rm X,bol} \geq 0.25  \times 10^{42} h^{-2}$erg s$^{-1}$.}
\label{alifig03}
\end{figure*}

In Fig.~\ref{alifig03}, we plot the magnitude gap
$\Delta m_{\rm 12}$ (within $0.5R_{\rm 200}$),
estimated for all 14628 X-ray bright groups (i.e. groups with $L_{\rm
X,bol} \geq 0.25 \times 10^{42} h^{-2}$erg s$^{-1}$) at $z=0$ as
a function of their mass fraction $\alpha_{1.0}$
at $z\!=\! 1$. 
Groups are colour-coded according to their dark matter
halo mass. 
The horizontal dashed line separates groups into fossils
($\Delta m_{\rm 12} \geq 2.0$) and non-fossils ($\Delta m_{\rm 12} <
2.0$). All data points on the right side of the {\it vertical
dashed-line} have assembled more than 50\% of their mass by $z \sim$1.0
and hence have a minimum assembly time of about $\sim$7.7 Gyr. The
contour lines represent the number of data points (groups) in each of
$25 \times 25$ cells of an overlaid grid which is equally spaced
along both the horizontal and vertical axes.
 
Three results emerge from this plot: (i) As is expected, on average
the rate of mass growth in massive systems is higher than in low mass
groups as the majority of massive groups and clusters have assembled
less than 50\% of their final mass at $z \sim$1.0. (ii) Less massive
groups (and therefore early-formed ones) tend to develop larger
magnitude gaps in comparison to massive groups and
clusters. Consequently the fraction of massive fossils, identified in
this way, is less than low mass fossil groups. (iii) For any given
$\alpha_{1.0} \gtrsim 0.5$, the majority of groups have magnitude gaps
$\Delta m_{\rm 12} \lesssim 2.0$, as is evident from the density of
contours. In other words, the number of  early-formed groups with values of
$\Delta m_{\rm 12} < 2.0$ exceeds the number of fossil groups with $\Delta
m_{\rm 12} \gtrsim 2.0$.

Unlike the first two results, the third conclusion is not in agreement
with our current view that early-formed groups necessarily develop
larger magnitude gaps.  Clearly, the majority of groups with similar
values of $\alpha_{1.0} \gtrsim 0.5$ have smaller magnitude gaps.
Without doubt, the parameter $\Delta m_{\rm 12}$ is influenced by the
infall and merging of galaxies and sub-groups within galaxy
groups. This could result in the increase (in case of merging) or
decrease (in case of the infalling of new galaxies) in $\Delta m_{\rm
  12}$. Indeed, in the work of \citet{vbb08}, one finds that the
``fossil'' phase of any fossil group is transient, since the magnitude
gap criterion will sooner or later be violated by a galaxy comparable
to the brightest galaxy falling into the core of the group.


\begin{figure*}
\epsfig{file=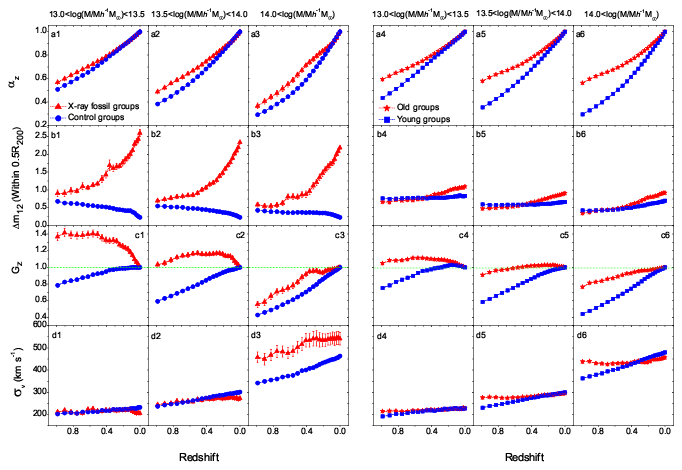,width=1.1\hsize}
\caption{ The evolution with redshift of various physical parameters
  of X-ray bright groups, in various ranges of group mass. {\it Left
    panel:} Haloes are classified as {\bf X-ray fossil} ($\Delta
  m_{\rm 12}\geq 2.0$, {\it red triangles}) and {\bf control }($\Delta
  m_{\rm 12}\leq 0.5$, {\it blue circles}) groups based on the
  magnitude gap between the first and the second brightest galaxies
  within $0.5R_{\rm 200}$. {\it Right panel:} Groups are divided into
  {\bf old}($\alpha_{1.0} \geq 0.5$, {\it red stars}) and {\bf young}
  ($\alpha_{1.0} \leq 0.5$, {\it blue squares}) population
  respectively. Each row represents the evolution of a parameter
  characteristic of galaxy groups that can be measured from
 the simulations (but not necessarily from observations). 
 From top to bottom the parameters on
  the y-axis represent: 
{\bf (i)}~$\alpha_z$, the ratio of the group halo
  mass at redshift $z$ to its mass at $z=0$ ($a1 ,..., a6$), {\bf
    (ii)}~$\Delta m_{\rm 12}$, i.e. the magnitude gap between the
  first two brightest group galaxies found within 0.5$R_{\rm 200}$
  ($b1 ,..., b6$), {\bf (iii)}~$G_z$, the ratio of the number of
  galaxies within $0.5\,R_{\rm 200}$ at redshift $z$ of a given galaxy
  group to the number of galaxies within $0.5\, R_{\rm 200}$ at redshift
  $z=0$ of the same group ($c1 ,..., c6$), and {\bf (iv)}~the group
  velocity dispersion $\sigma_V$ in km~s$^{-1}$ ($d1,...,d6$). 
   In the third panels from top, the
  horizontal {\it green dashed-lines} intersect 
   the y-axis at $G_z=1$.  }

\label{alifig04}
\end{figure*}

To quantify the above results, we study the evolution with redshift of
various physical parameters for two different sample of groups, drawn
from the distribution of galaxy groups in Fig.~\ref{alifig03}. In the
first sample, haloes are divided into {\bf old}($\alpha_{1.0} \geq
0.5$, $b+c$ in Fig.~\ref{alifig03}) and {\bf young} ($\alpha_{1.0} \leq
0.5$, $a+d$ in Fig.~\ref{alifig03}) groups respectively. In the second
population, haloes are classified as {\bf X-ray fossil} ($\Delta
m_{\rm 12}\geq 2.0$, $a+b$ in Fig.~\ref{alifig03}) and {\bf control
}($\Delta m_{\rm 12}\leq 0.5$, $c+d$ in Fig.~\ref{alifig03}) groups
based on the magnitude gap between the first and the second brightest
galaxies within half a virial radius of the centre of the group.

For each sample, the evolution of various parameters are shown
in two panels of Fig.~\ref{alifig04}.  From top to bottom these
parameters are:
\begin{itemize}

\item  $\alpha_z$, i.e. the ratio of the group halo mass at redshift $z$ to its mass at $z=0$,

\item  $\Delta m_{\rm 12}$ within 0.5$R_{\rm 200}$,

\item  Ratio of the number of galaxies within $0.5R_{\rm 200}$ at redshift $z$ of a given galaxy group to the number of galaxies within $0.5R_{\rm 200}$ at redshift $z=0$ of the same group, i.e.  $G_z$,


\item  Group velocity dispersion $\sigma_V$ in km~s$^{-1}$.

\end{itemize}

In each panel of Fig.~\ref{alifig04},
the left, middle, and right columns correspond to
different ranges in group mass, as indicated.  The left panel in
Fig.~\ref{alifig04} illustrates X-ray fossil ({\it red triangles}) and
control ({\it blue circles}) groups respectively while the right panel
shows old ({\it red stars}) and young ({\it blue squares}) groups. The
horizontal {\it green dashed-lines} intersect y-axes at
$G_z=1$. Errors on data points are the standard error on the mean,
i.e. $\sigma / \sqrt{N}$, where $\sigma$ is the standard deviation of
the original distribution and $N$ is the sample size.

A comparison between Figs.~\ref{alifig04}$a1,a2,a3$ and
Figs.~\ref{alifig04}$a4,a5,a6$ shows that older groups, which have been
picked up according to their lower rate of mass growth (i.e. larger
$\alpha_{1.0}$), represent a {\it perfect class} of fossils,
though they develop a magnitude gap $\Delta m_{\rm 12}$ which is not as
large as those seen in X-ray fossils (see also
Figs.~\ref{alifig04}$b1,...,b6$). 

On the other hand, unlike old
groups, X-ray fossils develop large magnitude gaps, 
which  do not necessarily correspond 
to their rapid mass growth, especially in massive groups with
$\log (M(R_{\rm 200})/h^{-1}\,$M$_{\odot}) \geq 14.0$. This reflects the
fact that the majority of real passive groups have a 
small magnitude gap
between their two brightest galaxies. Thus the expression $\Delta
m_{\rm 12} \geq 2$ only partially separates 
genuine old/passive groups
from young/forming groups, as there are a larger fraction of genuine
old groups but with small $\Delta m_{\rm 12}$.

From Figs.~\ref{alifig04}$c4,c5,c6$, it is obvious that older groups
are essentially more relaxed systems, which have not recently
experienced a major merger, as the rate of infall of galaxies is equal
or even less than the rate at which galaxies merge with the central
group galaxy. Therefore in old groups the parameter $G_z$ is more or
less constant with time, compared to that of the younger groups within
the same group mass bin.  The situation is a bit different in X-ray
fossils with $\log (M(R_{\rm 200})/h^{-1}\,$M$_{\odot}) \leq 14.0$
(Figs.~\ref{alifig04}$d1,d2$), since, in these cases, the rate of
galaxy merging is noticeably larger than that of the infall of
galaxies. As a result, very large magnitude gaps can develop in X-ray
fossil groups.  It is also evident both massive X-ray fossils and
control groups with $\log (M(R_{\rm 200})/h^{-1}\,$M$_{\odot}) \geq
14.0$ (Fig.~\ref{alifig04}$d3$) are in a state of rapid mass
growth. As a consequence, massive X-ray groups are not dynamically
relaxed systems as they are influenced by infall of galaxies and
substructures.

Finally, it is worth considering how the
velocity dispersion, plotted in Figs.~\ref{alifig04}$d1,...,d6$ changes 
with time in the
different kinds of groups. As Figs.~\ref{alifig04}$d4,d5,d6$ show, as
long as the rate of infall of subgroups is close to 1.0
({\it green dashed-line}), the velocity dispersion does not 
significantly change with time, which in turn is a sign that these groups are
certainly relaxed systems.



\begin{figure*}
\epsfig{file=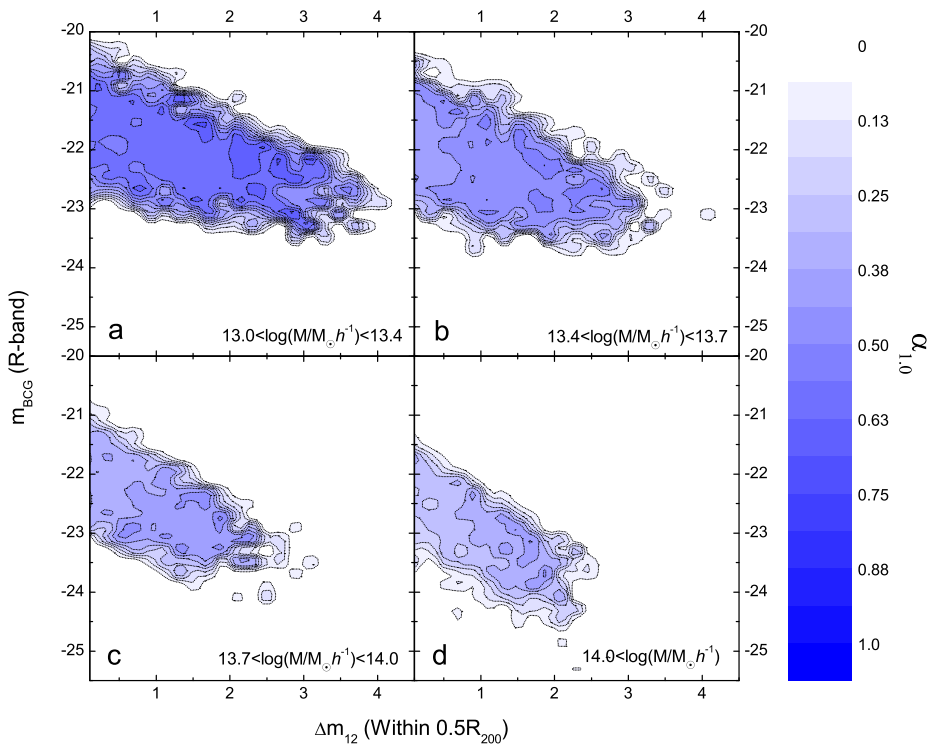,width=0.8\hsize}
\caption{ The absolute R-band magnitude of BCGs for all X-ray bright
  groups versus $\Delta m_{\rm 12}$ within 0.5R$_{\rm 200}$ in four
  different mass bins. A grid of $45\times 55$ cells has been
  superposed in each panel, and the cell is colour-coded according to
  the median value of $\alpha_{1.0}$ within the cell.  The contours
  trace the distribution of median $\alpha_{1.0}$ values in each
  panel. All panels have the same scale.}
\label{alifig05}
\end{figure*}

\subsubsection{BCG  magnitudes} 


Since the central galaxy in a fossil group is a product of numerous
mergers, many of them with luminosities close to L$_{\star}$ galaxies,
X-ray fossils are expected to be dominated by optically luminous
brightest galaxies (BCGs) more often than their non-fossil
counterparts. Here we explore the correlation between the luminosity
of the central galaxies of groups with large magnitude gaps, and their
mass assembly history.

The four panels in Fig.~\ref{alifig05} demonstrate the relation
between the absolute $R$-band magnitude of the BCGs for all X-ray bright
groups and the magnitude gap $\Delta m_{\rm 12}$ within $0.5\,R_{200}$ 
in four different mass bins. In each panel of
Fig.~\ref{alifig05}, a grid of $45\times 55$ cells is overlaid,
where each cell is colour-coded according to the median of
$\alpha_{1.0}$ in that cell. Accordingly, contours in each panel trace
the distribution of median $\alpha_{1.0}$ values.

Fig.~\ref{alifig05} shows clearly that both $\alpha_{1.0}$ and BCG
$R$-band magnitudes increase with decrease in group halo masses. But
it does not show a tight correlation between the R-band luminosity of
BCGs and group magnitude gaps $\Delta m_{\rm 12}$, though the correlation
is more pronounced in clusters with $M(R_{\rm 200}) \geq
10^{14}\,h^{-1}\,$M$_{\odot}$ (Fig.~\ref{alifig05}d).  
Therefore, imposing 
a magnitude cut for the  BCGs would result in the loss
from the sample 
of a large number of genuine groups 
which are not X-ray fossil systems
according to the optical condition involving $\Delta m_{\rm 12}$.

\subsection{The Fossil phase in the life of groups}
\label{Phase1}
 
The existence of large magnitude gaps in X-ray fossils in
Figs.~\ref{alifig04}$b1,b2,b3$ is expected as these groups were
initially selected according to their $\Delta m_{\rm 12}$ at $z$=0. It
would be interesting if they could be shown to have maintained such
large magnitude gaps for a longer time, in comparison to control
groups, which would be the case if X-ray fossils were relaxed groups
without recent major mergers. Also if fossil groups in general are the
end results of galaxy merging, then we do expect the majority of
fossils selected at higher redshifts to still be detected as fossils
at the present epoch. In other words, the {\it fossil phase} in
the life of a galaxy group should be long-lasting.

\begin{figure}
\epsfig{file=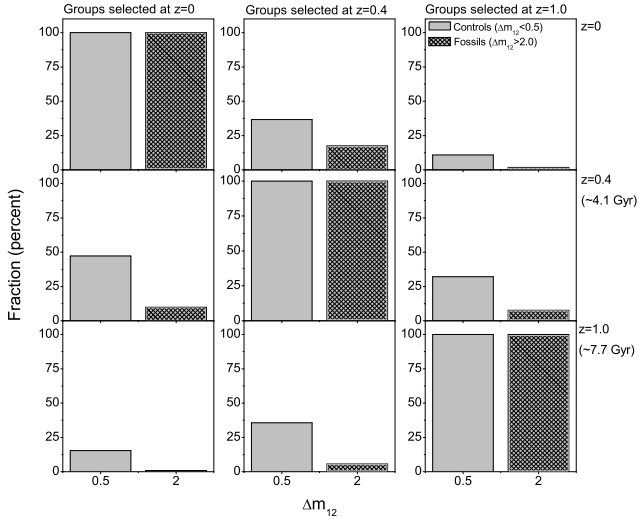,width=3.7in}
\caption{ The fate of fossil groups identified at different redshifts.
  Fossil groups ({\it dark shaded histogram}) and control
  groups ({\it grey shaded histogram}) are initially selected at
  $z$=0 ({\it left column}), $z$=0.4 ({\it middle column}), and
  $z$=1.0 ({\it right column}). For these objects, we explore
  what fraction remain ``fossil'' or ``control'' groups at two other
  epochs in the redshift range $0 \lesssim z \lesssim 1.0$.
  It is clear that the fossil phase does not last in $>90$\%
  of groups after 4~Gyr, no matter at which epoch they are identified.
 }
\label{alifig06}
\end{figure}

To test this, we select three sets of fossil groups with $\Delta
m_{\rm 12}\geq 2.0$ at three different redshifts.  By tracing the
magnitude gap of each set from $z$=1.0 to $z$=0, we examine the fossil
phase of each set in time. In Fig.~\ref{alifig06}, fossils ({\it black
  shaded histogram}) and control ({\it black thick line histogram})
groups are selected at $z$=0 ({\it left column}), $z$=0.4 ({\it middle
  column}), and $z$=1.0 ({\it right column}). Fractions of fossil and
control groups in each column of Fig.~\ref{alifig06} have been
separately estimated by normalising the number of fossil and control
groups at other redshifts to their total numbers at the redshift at
which they were initially selected.
 
This plot shows that, contrary to expectation, no matter at what
redshift the fossils are selected, after $\sim$4 Gyr, more than
$\sim$90\% of them change their status and become non-fossils
according to the magnitude gap criterion. Over the span of 7.7~Gyr,
which is the time interval between $z=$0-1, very few groups retain a
two-magnitude gap between the two brightest galaxies.  This means that
the fossil-phase is a temporary phase in the life of fossil groups
\citep[also see][]{vbb08}, and there is no guarantee that an
observed fossil group, at a relatively high redshift, remains a fossil
until the present time, if fossils are selected according to their
magnitude gap $\Delta m_{\rm 12}\geq 2.0$.

\section{Revising the Optical Criterion for finding fossil groups}
\label{RevOptDef}

Using the Millennium simulation DM runs as well as the gas and
semi-analytic galaxy catalogues based on them, it seems from the above
that the conventional optical condition $\Delta m_{\rm 12}\geq 2$,
used to classify groups as fossils, does not ensure that that these
systems represent a class of old galaxy groups, in which the central
galaxy has grown through the merging of other comparable group galaxy
members. Having said that, it is true that the magnitude gap in a
galaxy group is related to the mass assembly history of the group, for
we saw that in groups, such a gap develops gradually with
time. However, the difference between the luminosities of the two
brightest galaxies in groups is not always reliable for the
identification of fossil systems, as this quantity is vulnerable to
the assimilation of a comparable galaxy into the core of the group, as
a result of infall or merger with another group. 

We therefore attempt
to identify a more robust criterion, in terms of the difference of
optical magnitudes among the brightest galaxies in a group, which
might be better suited to identifying systems where most of the mass
has been assembled at an early epoch. 
As introduced in Sec.~\ref{evolution1}, we quantify the age of a group
in terms of the mass assembly parameter
$\alpha_{1.0}$, which for an individual group is
the ratio of its mass at redshift $z\!=\! 1.0$ 
to its present mass at $z\!=\! 0$,
i.e. $\alpha_{1.0}\equiv M_{z=1.0}/M_{z=0}$. We begin by considering the
effect of the radius within which the magnitude gap is calculated.

\subsection{A general criterion for the magnitude gap}

Assume a general optical condition in defining early-formed groups
according to the magnitude gap between the brightest group galaxy and
other group members in the following form:

\begin{equation}
\Delta m_{1i} \geq j,
\label{dm1i}
\end{equation}
where $\Delta m_{1i}$ is the difference in $R$-band magnitude between
the first brightest group galaxy and the $i^{th}$ brightest group
galaxy within 0.5R$_{\rm 200}$ (or R$_{\rm 200}$) of the group centre.
The current definition of fossils involves $i=2$ and $j=2$. Obviously
any group satisfying Eq.~\ref{dm1i} must contain at least $i$
galaxies.  We do not consider $i>10$ since then we have to exclude
most groups in our sample, and it would turn out not be very useful
for observers as well.

 
\begin{figure*}
\centering
\epsfig{file=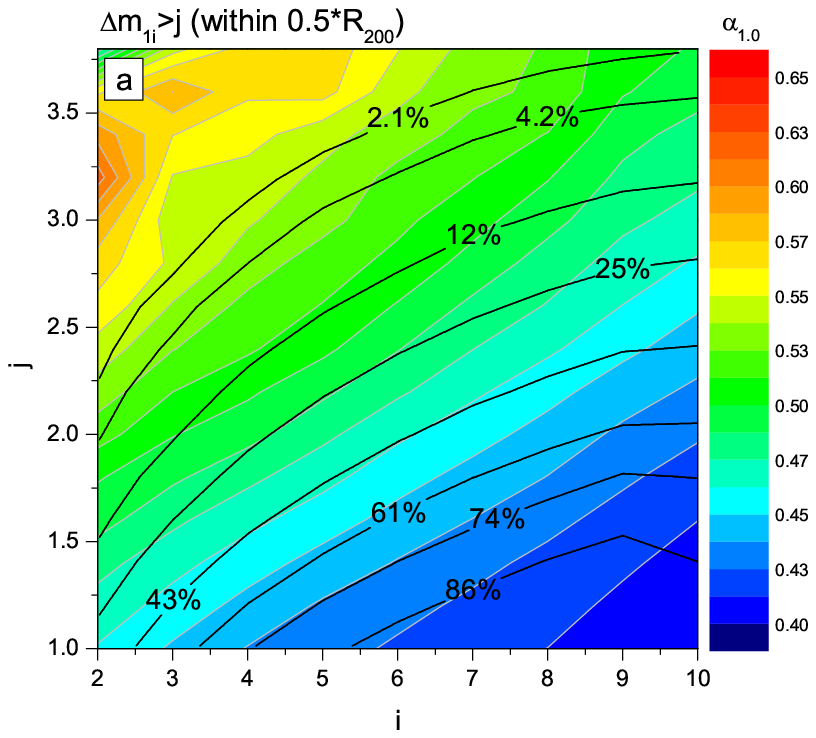,bb=50 20 290 250,clip=,width=0.49\hsize}
\epsfig{file=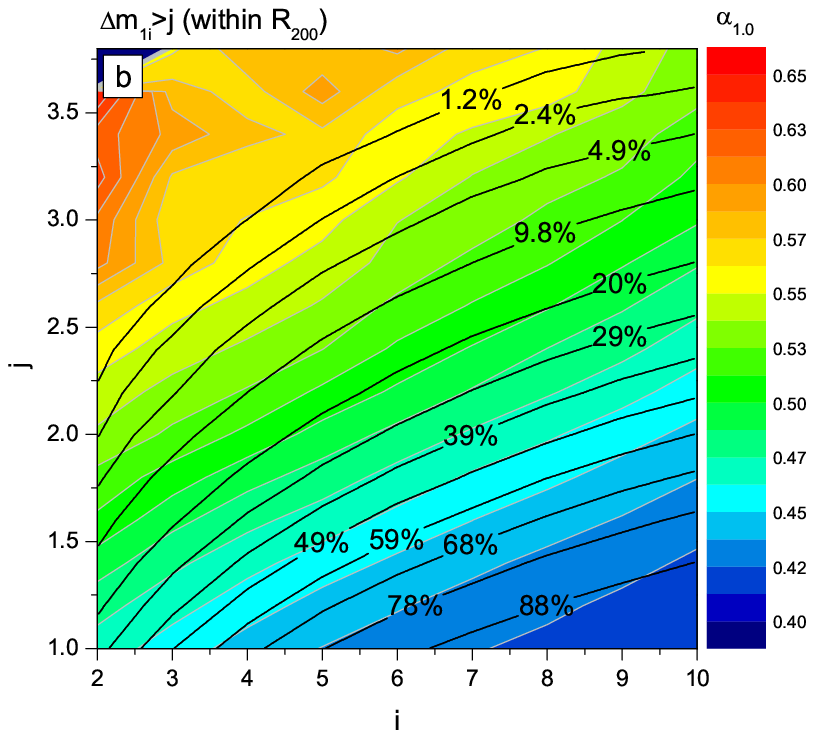,bb=50 20 290 250,clip=,width=0.49\hsize}
\caption{The dependence on ($i$, $j$), defined in Eq.~\ref{dm1i}, of
  the mass assembly parameter $\alpha_{1.0}$, which is defined as the
  ratio of the mass of a group at redshift $z\!=\! 1.0$ to its mass at
  $z\!=\! 0$, i.e. $\alpha_{1.0}\equiv M_{z=1.0}/M_{z=0}$.  For each
  value of $i$, groups are sorted according to the value of their
  magnitude gaps $\Delta m_{1i}$ calculated within a certain radius
  (different for the two panels). Then, for different values of $j$,
  the average for $\alpha_{\rm 1.0}$ is calculated for all groups
  satisfying Eq.~\ref{dm1i}.  The plot is colour-coded according to
  $\alpha_{\rm 1.0}$. The {\it black contours} are drawn such that the fraction
of the total number of groups identified is constant along each line.
The magnitude gap $\Delta m_{1i}$
  is calculated  {\bf (Panel a.)} 
within half the overdensity radius, i.e. 0.5R$_{\rm
    200}$, and  {\bf (Panel b.)} 
  within the overdensity radius, i.e. R$_{\rm 200}$.}
\label{alifig07}
\end{figure*}

As we consider the magnitude gap between the brightest to the 
$i^{\rm th}$(=2, 3,..., 9, 10) brightest galaxy, 
the value of the magnitude gap varies from $j\gtrsim$0 to
$j\lesssim$5. Our aim is to find a pair  of ($i$, $j$) in  
Eq.~\ref{dm1i} which yields the best selection 
of genuinely old groups with a history of 
early mass assembly. 

In Fig.~\ref{alifig07}, we show how the parameter $\alpha_{1.0}\equiv
M_{z=1.0}/M_{z=0}$, which represents the mass assembly of groups since
redshift $z$=1.0, depends upon the selection of $i$ and $j$ in
Eq.~\ref{dm1i}. For each value of $i$ in Fig.~\ref{alifig07}, groups
are first sorted according to their magnitude gaps $\Delta m_{1i}$
estimated within 0.5R$_{\rm 200}$ or R$_{\rm 200}$), where the latter
is the overdensity radius of the group. For each $i$, the average
value for $\alpha_{\rm 1.0}$ is calculated, for each value of $j$,
for all groups satisfying Eq.~\ref{dm1i}. The plot is colour-coded
according to the values of $\alpha_{\rm 1.0}$. The {\it black contours}
give an idea of the number of groups involved:
the fraction of the total number of groups
identified by parameters ($i$, $j$) is constant along each of these lines.

From Fig.~\ref{alifig07}a, for instance, we find that systems with
$i=4$ and $j=3$, i.e. systems for which $\Delta m_{14} \geq 3$ (within
0.5R$_{\rm 200}$), yield $\sim$2.4$\%$ of groups with $\alpha_{\rm
  1.0} \sim$0.54.  The same is $\sim$1.2$\%$ with $\alpha_{\rm 1.0}
\sim$0.56, if one estimates $\Delta m_{14} \geq 3$ within R$_{\rm
  200}$ according to Fig.~\ref{alifig07}b.

In fact, by changing our search radius from 0.5R$_{\rm 200}$ to
R$_{\rm 200}$, we find $\sim$50$\%$ fewer groups satisfying
Eq.~\ref{dm1i}, for the same value of $\alpha_{\rm 1.0}$.  Therefore,
hereafter, we just use Panel~{\bf a} of Fig.~\ref{alifig07} which
estimates the magnitude gap $\Delta m_{1i}$ within half of the
overdensity radius, i.e. 0.5R$_{\rm 200}$. From this plot, the
fraction of groups picked out by applying the conventional fossil
criterion $\Delta m_{\rm 12} \geq 2$ is $\sim$4.0$\%$ with
$\alpha_{\rm 1.0} \sim$0.52.

If we were to find an improved criterion for
finding fossils, a better set of parameters ($i$, $j$) in
Eq.~\ref{dm1i} should 
\begin{enumerate}
\item identify groups with larger
valuer of $\alpha_{\rm 1.0}$, and/or 
\item find a larger fraction
of groups with the same or larger value of $\alpha_{\rm 1.0}$, 
\end{enumerate}
than found
in conventional fossils, i.e. groups with  ($i$, $j$)=(2, 2).

For example, by choosing $i\!=\! 4$ and $j\!=\! 3$, the fraction of
groups found with $\Delta m_{\rm 14} \geq 3$ turns out to be
$\sim$40$\%$ less than when $i$=$j$=2, but it would identify slightly
older groups, with an average $\alpha_{\rm 1.0} \sim$0.55, whereas the
average $\alpha_{\rm 1.0} \sim$0.52 in fossils with $i$=$j$=2.  In
other words, ($i$, $j$)=(4, 3) identifies marginally older groups, at
the expense of losing a large number of early-formed groups, compared to 
 the case of conventional fossils ($i$, $j$)=(2, 2).

 Exploring Fig.~\ref{alifig07}a, we adopt ($i$, $j$)=(4, 2.5) as an
 example of how the fossil search criterion can be improved.  If we
 define all groups with $\Delta m_{\rm 14} \geq 2.5$ within
 $0.5\,R_{\rm 200}$ as fossils, then the we would find groups with
 on average the same mass assembly history, i.e, the same average
 value of $\alpha_{\rm 1.0}$, but we would identify $\sim$50$\%$ more
 such groups, compared to groups identified the conventional parameters
 ($i$, $j$)=(2, 2).

We will examine such groups further in the next section. Meanwhile,
the Figs.~\ref{alifig07} will allow the user to find
their favourite combination of ($i$, $j$) for both 
0.5R$_{\rm 200}$ and R$_{\rm 200}$.


\begin{table*}
\begin{minipage}{120mm}
\centering
\caption{Peak values (from Gaussian fits) of the histograms of the
  mass assembly parameter $\alpha_{1.0}$ for various classes of
  groups.  (see Fig.~\ref{alifig08}).  Group halo mass $M(R_{\rm
    200})$ is in units of $\,h^{-1}\,$M$_{\odot}$.  F$_{\rm 14}$
  consists of all groups with $\Delta m_{\rm 14} \geq 2.5$, and
  F$_{\rm 12}$, those with $\Delta m_{\rm 12} \geq 2.0$, both within
  0.5R$_{200}$ of the group centre. }
  \begin{tabular}{lccc}
    \\ \hline 
    Group type 
  &   $ 13.0 \leq \log M(R_{\rm 200}) $    & $13.0 \leq \log M(R_{\rm 200}) \leq 13.5$ & $\log M(R_{\rm 200}) \geq 13.5$    \\  
    & Panel {\bf a} &    Panel {\bf b} & Panel {\bf c}                 \\  \hline
    All X-ray bright groups                           &   $0.41 \pm 0.01$          & $0.52\pm0.01$	 & $0.38\pm0.01$ \\
    F$_{\rm 12}$                                  &   $0.53 \pm 0.01$          & $0.56\pm0.01$   & $0.49\pm0.01$ \\
    F$_{\rm 14}$                                  &   $0.52 \pm 0.01$          & $0.55\pm0.01$	 & $0.48\pm0.01$ \\
    F$_{\rm 12}$ $\cap$ F$_{\rm 14}$              &   $0.54 \pm 0.01$          & $0.55\pm0.01$	 & $0.50\pm0.02$ \\\hline \\
\end{tabular}
\label{fitPARAM}
\end{minipage}
\end{table*}

\begin{figure*}
\epsfig{file=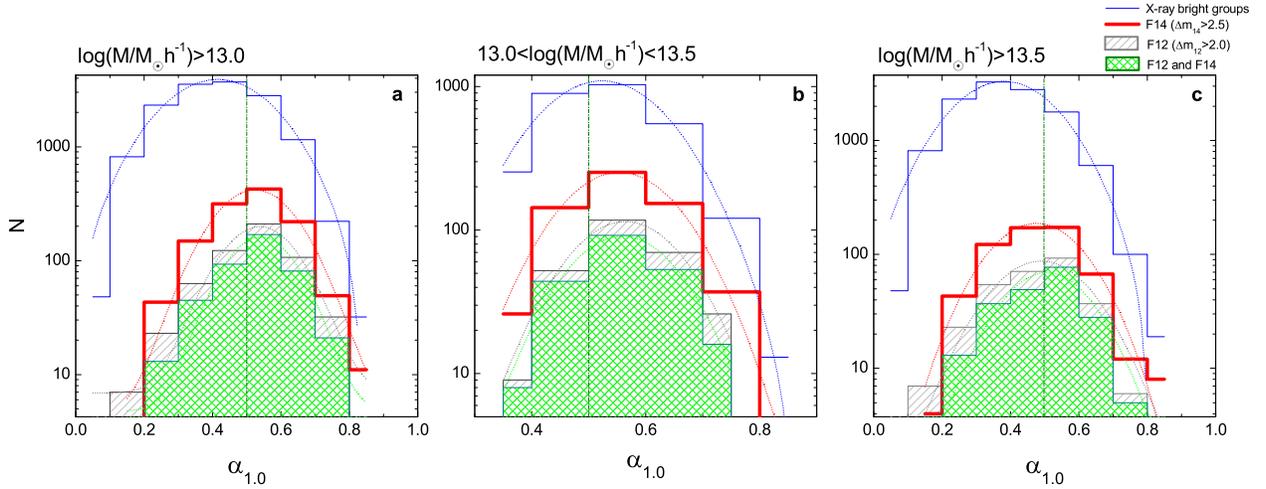,width=7.25in}
\caption{Histograms of the mass assembly parameter $\alpha_{\rm 1.0}$
  for X-ray bright groups ({\it blue histogram}), groups that satisfy
  the criterion F$_{\rm 14}$ ({\it red thick histogram}), those that
  satisfy F$_{\rm 12}$ ({\it gray shaded histogram}), and groups that
  satisfy both criteria, i.e.  F$_{\rm 12}$ $\cap$ F$_{\rm 14}$ ({\it
    green shaded histogram}). Overlaid are Gaussian fits to each
  histogram (see Table~\ref{fitPARAM}). Panels {\bf a}, {\bf b}, and
  {\bf c} correspond to the logarithm of the
  group halo mass in the range $ 13.0 \leq \log
  M(R_{\rm 200}) $, $13.0 \leq \log M(R_{\rm 200}) \leq 13.5$, and
  $\log M(R_{\rm 200}) \geq 13.5$ respectively, the unit of
  $M(R_{\rm 200})$ being $\,h^{-1}\,$M$_{\odot}$. The {\it
    green dash dotted} line intersects the $x$-axis at $\alpha_{\rm
    1.0}=0.5$ where haloes have assembled 50$\%$ of their mass at
  redshift $z$=1.0.}
\label{alifig08}
\end{figure*}

\subsection{The optical criterion $\Delta m_{\rm 14} \geq 2.5$ within
  0.5R$\, _{200}$}
\label{altcrit}

Having explored alternative criteria for identifying groups with a
history of early formation, we now compare the history of mass
assembly of groups selected according to $\Delta m_{\rm 14} \geq 2.5$
(these groups are hereafter collectively referred to as {\it F$_{\rm
    14}$}) with those groups selected according to $\Delta m_{\rm 12}
\geq 2.0$ (hereafter {\it F$_{\rm 12}$}), both within 0.5R$_{200}$ of
the group centre.  The latter category are the conventional fossil
groups.

The {\it blue histogram} in Fig.~\ref{alifig08} represents the
distribution of the mass assembly parameter $\alpha_{\rm 1.0}$ for all
X-ray bright groups (as defined in Fig.~\ref{alifig01}) in our sample.
It also shows the groups in the categories F$_{\rm 14}$ ({\it red
thick histogram}) and F$_{\rm 12}$ ({\it grey shaded histogram}). 
The {\it green shaded histogram} corresponds to groups
that satisfy both criteria, i.e.  F$_{\rm 12}$ $\cap$ F$_{\rm 14}$.
The {\it green dash dotted} line intersects the x-axis at $\alpha_{\rm
  1.0}=0.5$, representing groups for which half of their mass had been
assembled at redshift $z\!=\! 1$.  Gaussian fits to each histogram are
overlaid.

Panels {\bf a}, {\bf b}, and {\bf c} in Fig.~\ref{alifig08} correspond
to different 
ranges of the logarithm of
the group halo mass $ 13.0 \leq \log M\,(R_{\rm 200}) $,
$13.0 \leq \log M\,(R_{\rm 200}) \leq 13.5$, and $\log M\,(R_{\rm
  200}) \geq 13.5$ respectively, where  $M\,(R_{\rm
  200})$ is in units of $\,h^{-1}\,$M$_{\odot}$. These figures, as
well as the values of the peaks of Gaussian fits to the distribution
of $\alpha_{\rm 1.0}$ in each case (given in Table~\ref{fitPARAM})
lead to the following observations:

\begin{enumerate}
\item The groups belonging to F$_{\rm 12}$ and
F$_{\rm 14}$ are older than the overall population of X-ray bright
groups (for all values of halo mass), 
though such a difference is less pronounced
in low-mass systems. Since haloes are thought
to be hierarchically assembled,
one expects to find a higher incidences of early-formed and low-mass
groups in comparison to massive systems.

\item Within the errors, groups belonging to F$_{\rm 14}$ are
  almost as old as F$_{\rm 12}$, i.e. the estimated $\alpha_{\rm 1.0}$
  in F$_{\rm 12}$ systems (as given in Table. ~\ref{fitPARAM}) is more
  or less the same as found in F$_{\rm 14}$ systems.  However, the
  fraction of groups in category
  F$_{\rm 14}$ is at least 50$\%$ more than F$_{\rm 12}$.
  This tells us that in general, the criterion 
  $\Delta m_{\rm 14} \geq 2.5$ has a
  higher efficiency of identifying early-formed systems than $\Delta
  m_{\rm 12} \geq 2.0$.

\item Interestingly, $\sim 75\%$ of F$_{\rm 12}$ haloes in
  Fig.~\ref{alifig08}{\bf a} also fulfil the $\Delta m_{\rm 14} \geq
  2.5$ condition. Conversely, $\sim 35\%$ of F$_{\rm 14}$ haloes
  satisfy the $\Delta m_{\rm 12} \geq 2.0$ criterion. This means that
  a large proportion of the population of early-formed groups in the
  category F$_{\rm 14}$ is different from those in F$_{\rm 12}$.
  Groups which satisfy both criteria, i.e.  F$_{\rm 12}$ $\cap$
  F$_{\rm 14}$, are not necessarily older in comparison to those
  belonging to either F$_{\rm 12}$ or F$_{\rm 14}$ (see
  Table. ~\ref{fitPARAM}).

\item Fig.~\ref{alifig08}{\bf b} shows that
in fact neither criterion $\Delta m_{\rm 12} \geq 2.0$ nor $\Delta
m_{\rm 14} \geq 2.5$ is efficient in finding early-formed
groups in low-mass regime, even among X-ray bright groups. 

\end{enumerate}

In the following section, we compare the environment and 
abundance of the groups belonging to the F$_{\rm
  12}$ and F$_{\rm 14}$ categories.

\subsection{The local environment of fossil groups}
\label{Delta4}


If galaxy mergers are responsible for the absence of bright galaxies
in groups such as X-ray bright fossils, then most of the matter infall
into these systems would have happened at a relatively earlier epoch.
Consequently, at the present time, old groups should be more isolated
than groups which have recently formed \citep[e.g.][]{labarb09}. Here we
 examine the local environment of groups, using
the density
parameter $\Delta_{4}$, defined as the number of haloes within a
distance of $4\,h^{-1}$~Mpc from the centre of each group. The
the local densities are calculated at $z=0$ according to
\begin{equation}
\Delta_4=\frac{\rho_4}{\rho_{bg}}-1,
\label{D4}
\end{equation}
where $\rho_{4}$ is the number density of haloes within a spherical
volume of of 4$h^{-1}$~Mpc in radius, and $\rho_{bg}$ is the
background density of haloes within the whole volume of the Millennium
simulation.  Since the mass assembly of groups is mostly influenced by
the infall of subgroups, which individually have masses typically
below $\sim$10\% (and often substantially smaller) of the parent halo
mass, it is important to take into account all haloes with $M\,(R_{\rm
  200}) \geq 10^{11}\, h^{-1}\,$M$_{\odot}$ from the FoF group
catalogue in order to estimate $\Delta_4$.

From Gaussian fits to the histograms of the local density
$\Delta_4$ of F$_{\rm 12}$ and F$_{\rm 14}$, control groups, as well
as those of all X-ray bright groups, we find
\begin{equation}
\Delta_4  = \left\{
\begin{array}{rll}
6.31 \pm 0.17 & {\rm for } & \mbox{Control groups ($\Delta m_{\rm 12}\leq 0.5$)}\\
6.25 \pm 0.18 & {\rm for } & \mbox{X-ray bright groups}\\
5.10 \pm 0.35 & {\rm for } & {\rm F_{\rm 12}}\\
5.19 \pm 0.26 & {\rm for } & {\rm F_{\rm 14}}
\end{array} \right.
\end{equation}
where $\Delta_4$ is estimated using Eq.~\ref{D4}. It seems that both
F$_{\rm 12}$ and F$_{\rm 14}$ groups are more likely to lie in lower
density regions than control groups and X-ray bright groups.  This is
in agreement with our expectation as early-formed groups are assumed
to be in low-density local environments. However, the local density
around F$_{\rm 12}$ and F$_{\rm 14}$ groups is more or less the
same. This is consistent with the fact that F$_{\rm 12}$ and F$_{\rm
  14}$ have similar values of $\alpha_{\rm 1.0}$ (see
Table~\ref{fitPARAM}).

A similar trend is seen for density measures $\Delta_{5}$ and
$\Delta_{6}$, but as the sampling volume increases ($> \Delta_{6}$),
the above trend disappears, showing that the trend is related to the
immediate environment of groups.

%

\subsection{The abundance of fossil groups}
\label{abundance}

Various studies have shown that the fraction of early-formed groups
increases as the group halo mass decreases
\citep[e.g.][]{Milos06,Dariush07}. This phenomenon reflects the fact
that structures form hierarchically, where small virialised groups
form early, whereas most massive clusters form late. As the merging of
galaxies in clusters is less efficient than in groups, due to the high
velocity dispersion of cluster galaxies, clusters are less likely to
develop large magnitude gaps. At the same time, in low-mass groups
\citep[see,e.g.,][]{miles04,miles06} dynamical friction is more
effective in ensuring galaxies fall to the core of the system, due to
the smaller relative velocities involved.  As a result, the existence
of large magnitude gaps should be more frequently found in groups than
in clusters. Thus, to find an old population of groups according to
some criterion, and to study the way the criterion depends on group
halo mass, would be a good test for the validity of the condition.

\begin{figure}
\epsfig{file=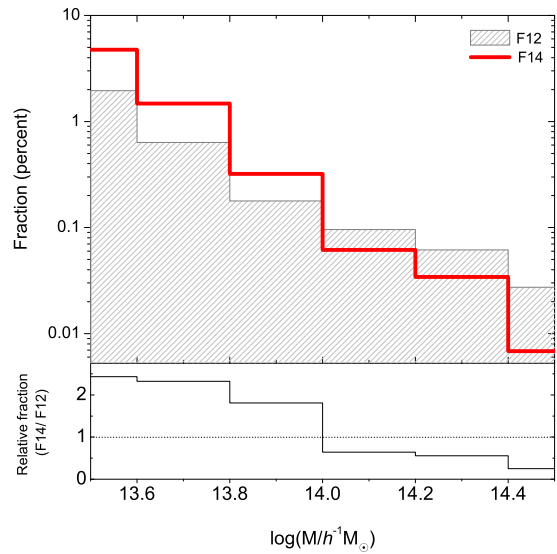,width=3.5in}
\caption{{\bf Top panel:} The abundance of F$_{\rm 12}$({\it gray
    shaded histogram}) and F$_{\rm 14}$({\it red thick line}) groups,
  i.e., the fraction of groups in each category as a function of halo
  mass.  {\bf Bottom panel:} The relative fraction of F$_{\rm 14}$
  over F$_{\rm 12}$ groups as a function of halo mass.}
\label{alifig09}
\end{figure}

The {\it top panel} of Fig.~\ref{alifig09} displays the abundance of
F$_{\rm 12}$ ({\it gray shaded histogram}) and F$_{\rm 14}$({\it red
  thick line}) groups, defined as the fraction of haloes in each
category, as a function of halo mass. The range of halo mass explored
is $\log M(R_{\rm 200}) \gtrsim 13.4$ in units of
$\,h^{-1}\,$M$_{\odot}$. Below this mass limit, the number of groups
abruptly decrease, since all groups here have been chosen to be X-ray
bright groups (see Fig.~\ref{alifig01}).

The plot shows that in comparison to the F$_{\rm 12}$ groups, the
F$_{\rm 14}$ groups are populated by less massive haloes.  This can be
better seen in the {\it bottom panel} of Fig.~\ref{alifig09}, where
the relative fraction of F$_{\rm 14}$ groups over F$_{\rm 12}$ is
shown. It can be inferred that, on average, the fraction of F$_{\rm
  14}$ groups with halo mass $M(R_{\rm 200}) \leq
10^{14}\,h^{-1}\,$M$_{\odot}$ is at least 50$\%$ more than the
fraction of F$_{\rm 12}$ . However, in the mass range $M(R_{\rm 200})
\geq 10^{14}\,h^{-1}\,$M$_{\odot}$, the fraction of F$_{\rm 14}$
groups decreases, though since the overall numbers in the extremely
high mass range are low, the statistics are poorer.


\subsection{The survival of the magnitude gap:  F$_{\rm 12}$ vs. F$_{\rm 14}$}
\label{FossilPhase2}

In Sec.~\ref{Phase1}, it was found that, in general, for the
conventional fossil groups (F$_{\rm 12}$), the {\it fossil phase} is
transient, $\gtrsim 90$\% of such groups ceasing to remain fossils after
4~Gyr.  Here we examine whether the fossil phase in $F_{14}$ groups
fares better.

\begin{figure}
  \epsfig{file=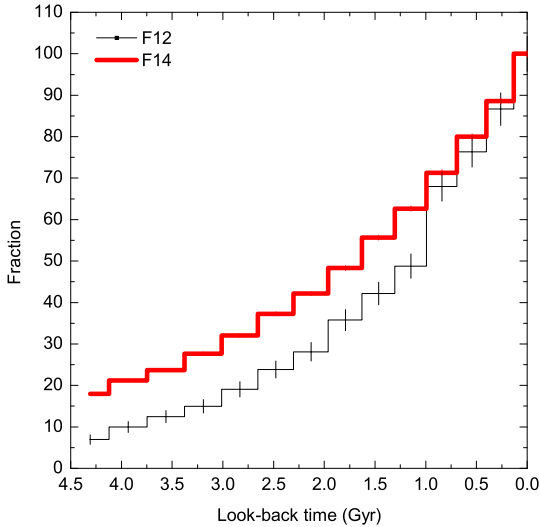,width=3.5in}
\caption{The fractions of F$_{\rm 12}$ ({\it black line}) and F$_{\rm 14}$ ({\it thick red line}) groups, identified at redshift $z\!=\! 0$, that survive
as fossils, as a function of look-back time.}
\label{alifig10}
\end{figure}

The histograms in Fig.~\ref{alifig10} represent the fractions of
F$_{\rm 12}$ ({\it black line}) and F$_{\rm 14}$ ({\it thick red
  line}), as a function of look-back time in Gyr.  The plot shows that
in comparison to the F$_{\rm 12}$ groups, the {\it fossil phase} lasts
longer by almost 1~Gyr for the same fraction of F$_{\rm 14}$ groups.
For example, the fraction of F$_{\rm 12}$ groups that maintains its
magnitude gap after $\sim$2.2 Gyr, falls to 28$\%$, while the
corresponding period is $\sim$3.2 Gyr in F$_{\rm 14}$.  Thus, not only
does the $\Delta m_{\rm 14} \geq 2.5$ condition identify at least 1.5
times as many fossil groups as the F$_{\rm 12}$ condition, it also
identifies groups in which the fossil phase lasts significantly
longer.  This can be explained from our analysis of the halo mass
distribution within F$_{\rm 12}$ and F$_{\rm 14}$, already discussed
in Sec.~\ref{abundance}.

\subsection{Comparison with observed groups}
\label{comparison}

When making detailed comparisons between simulations and
  catalogues of galaxies and groups compiled from observations, one
  has to be aware that simulated dark matter haloes have limited
  resolution, as mentioned in \S~\ref{SAMcroton} and ~\ref{SAMbower}.
  This means that all galaxies in the semi-analytic models, assigned
  to a particular halo, might not belong to dark matter subhalos.  We
  find that even after applying a magnitude cut, there would be a
  significant number of modelled galaxies, whose orbits are
  analytically calculated, that would end up being not a member of a
  sub-halo, and thus would not be classified as a group member.

  While dealing with magnitude gaps $\Delta m_{12}$ and $\Delta
  m_{14}$, it is worth examining to what extent these quantities are
  vulnerable to resolution effects like the above. A direct way would
  be to compare the magnitude gap distribution of groups, selected based on
  their $\Delta m_{\rm 14}$, between simulations and
  observations. This is not a straightforward task, as 
  groups identified from observational sky surveys are biased due to
  incompleteness in measured magnitude and redshift. Furthermore,
  a variety of  group finding algorithms are been adopted to identify
  groups in simulations and observation, which adds uncertainties to any
  such comparison.

  Here, we use the group catalogue of \citet{Yang07}, which uses a
  halo-based group finder on the Sloan Digital Sky Survey (SDSS
  DR4). They define groups as systems whose dark matter haloes, have
  an overdensity of 180, determined from dynamics. This makes this
  catalogue suitable for comparison with the Millennium simulation,
  where dark matter halos have an overdensity of 200. From Sample~II
  of the catalogue, groups with following properties are selected:
\begin{itemize}
 \item
  they have at least four members, 
\item they 
are within
  the redshift range $0.01 \leq z \leq 0.1$, and
\item  their estimated halo mass
  is $\log M\,(R_{\rm 180}) \geq 13.25, h^{-1}\,$M$_{\odot}$, since
  our Millennium X-ray groups have a similar
   mass threshold (see Fig.~\ref{alifig01}).

\end{itemize}

\begin{figure}
  \epsfig{file=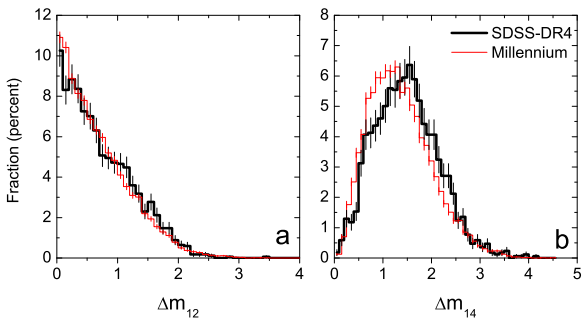,width=3.5in}
  \caption{The $R$-band magnitude gap distribution for haloes from the
    Millennium semi-analytic models of \citet{Bower06} ({\it red
      histograms}) superposed on the $r$-band data from Sample~II of
    SDSS-DR4 group catalogue of \citet{Yang07} ({\it black
      histograms}).  (a) This shows the magnitude gap $\Delta m_{\rm
      12}$ between the the first and second most luminous galaxies,
    compared with galaxies from the SDSS-DR4 catalogue of groups
    computed within group radius.  (b) This is the same as in (a), but
    for the magnitude gap $\Delta m_{14}$ between the first and the
    fourth most luminous galaxies. The 1697 SDSS-DR4 groups are within
    the mass range $\log (M(R_{\rm 180})/h^{-1}\,$M$_{\odot}) \geq
    13.25$, and redshift range $0.01 \leq z \leq 0.1$. Those from
    the Millennium simulations 
    consist of 14612 X-ray groups selected at redshift
    $z \sim 0.04$ in the same mass range.}
\label{alifig11}
\end{figure}

After applying the above criteria, 1697 groups were identified, and their
magnitude gaps were compared with galaxy groups selected from the
Millennium simulation at redshift $z \sim 0.041$. The magnitude gap
distributions from both SDSS-DR4 as well as the Millennium simulation
are shown in Fig.~\ref{alifig11}. Panels {\bf a} and {\bf b} of
Fig.~\ref{alifig11} refer to the magnitude gap distribution $\Delta
m_{\rm 12}$ and $\Delta m_{\rm 14}$ respectively. Results show that
the estimated magnitude gaps are in fair agreement with
observation. The fraction of groups with $\Delta m_{\rm 12} \geq 2.0$
is more or less the same while the fraction of those galaxy groups
with $\Delta m_{\rm 14} \geq 2.5$ is different by $\sim 1\%$ (see
Table.~\ref{sdssMill}). This shows that the incompleteness resulting from
limited resolution does not affect our statistics.

\begin{table}
\centering
\caption{Comparison between the observed and simulated fraction of groups 
with magnitude gaps $\Delta m_{\rm 12}$ and $\Delta m_{\rm 14}$, 
estimated from the histograms presented in Fig.~\ref{alifig11}.}
  \begin{tabular}{lll}
    \\ \hline 
    Selection criterion   &   SDSS (DR4)     & Millennium simulation   \\  
           &   \citep{Yang07}    & \citep{Bower06}    \\  \hline
    $\Delta m_{\rm 12} \geq$2.0 &   $2.0\% \pm 0.4$          & $2.1\%\pm0.2$ \\
    $\Delta m_{\rm 14} \geq$2.5 &   $6.2\% \pm 0.6$          & $5.1\%\pm0.2$ \\\hline \\
\end{tabular}
\label{sdssMill}
\end{table}

\section{Discussion and conclusions}%
\label{Discussion}

In this work, we analysed the evolution of the magnitude gap (the
difference in magnitude of the brightest and the $n$th brightest
galaxies) in galaxy groups.  Using the Millennium dark matter
simulations and associated semi-analytical galaxy catalogues and gas
simulations, we investigated how the magnitude gap statistics are
related to the history of mass assembly of the group, assessing
whether its use as an age indicator is justified.

A catalogue of galaxy groups, compiled from the Millennium dark matter
simulations, was cross-correlated with catalogues resulting from hot
gas simulations, and from semi-analytic galaxy evolution models based
on these simulations. This resulted in a list of groups, with various
properties of the associated dark matter haloes and galaxies, at 21
time steps, over the redshift range $z \simeq 1.0$ to $z\! =\! 0$.
The simulated X-ray emitting hot IGM properties were known for these
haloes only for $z\! =\! 0$, and these were used to define a sample of
X-ray emitting groups. This is necessary since our objective was to
examine the evolution of fossil groups, which are observationally
defined in terms of both optical and X-ray parameters.

We compared the estimated magnitude gaps in these galaxy groups from
two different semi-analytic models of \citet{Bower06} and
\citet{Croton06}, based on the Millennium dark matter simulations, and
found that the model of \citet{Bower06} better matches the observed
present-day distribution of the difference in magnitude between the
brightest galaxy in each group, and the second and third brightest
galaxies $\Delta m_{12}$ and $\Delta m_{13}$. We decided to use the
\citet{Bower06} catalogue for the rest of this study.

We examined the evolution with time of fossil galaxy groups,
conventionally defined as those with an $R$-band difference in
magnitude between the two brightest galaxies $\Delta m_{\rm
  12}\!\geq\! 2$ (within $0.5 \, R_{\rm 200}$ of the group centre). We
explored the nature of the groups that would be selected if the radius
of the group were extended to $R_{\rm 200}$, and the definition of the
magnitude gap in terms of $\Delta m_{\rm 1i}$ were varied.
Our major conclusions from the analyses can be summarised as follows:

\begin{enumerate}

\item The parameter $\Delta m_{1i}$ defined for a galaxy system as the
  magnitude gap between the first and $i^{th}$ brightest galaxies
  (estimated within a radius of 0.5R$_{\rm 200}$ or R$_{\rm 200}$) can
  be shown to be linked to the halo mass assembly of the system
  $\alpha_{\rm 1.0}$ (Fig.~\ref{alifig07}), such that {\it galaxy
    systems with larger magnitude gaps $\Delta m_{1i}$ are more likely
    to be early-formed than those with smaller magnitude gaps}.

\item Fig.~\ref{alifig06} shows that, contrary to expectation,
  irrespective of the redshift at which fossil groups are identified
  according to the usual criteria, after $\sim$4 Gyr, more than
  $\sim$90\% of them become non-fossils according to the magnitude gap
  criterion. Over the span of 7.7~Gyr, which is the time interval
  between $z=$0-1, very few groups retain a two-magnitude gap between
  the two brightest galaxies.  This provides clear evidence that the
   fossil phase is a
  temporary phase in the life of fossil groups \citep[also
  see][]{vbb08}.

\item In a given galaxy group, the merging of the $i^{th}$ brightest
  (or a brighter) galaxy, with the brightest galaxy in the group
  (often the central galaxy if there is one), results in an increase
  of $\Delta m_{1i}$. However, one of the main reasons for the fossil
  phase to be a transient one is that such a magnitude gap could be
  filled by the infall of equally massive galaxies into the core of
  the group, which would lead to a decrease in $\Delta
  m_{1i}$. Therefore, groups with smaller magnitude gaps are not
  necessarily late-formed systems.  Many groups spend a part of their
  life in such a fossil phase, though an overwhelming majority of them
  would not fulfil the criteria of the ``fossil'' label at all
  epochs.

\item For our sample of X-ray bright groups, the optical criterion
  $\Delta m_{\rm 14} \geq 2.5$ in the $R$-band is more efficient in
  identifying early-formed groups than the condition $\Delta m_{\rm
    12} \geq 2.0$ (for the same filter), and is shown to identify at
  least $50\%$ more early-formed groups. Furthermore, for the groups
  selected by the latter criterion, the {\it fossil phase} in general
  is seen on average to last $\sim 1.0~$Gyr more than their
  counterparts selected using the conventional criterion.

\item  Groups selected according to $\Delta m_{\rm 14} \geq 2.5$ at $z=0$
  correspond to $\sim 75\%$ of those identified using the $\Delta m_{\rm 12}
  \geq 2.0$ criterion. On
  comparing different panels in Fig.~\ref{alifig08}, one finds that
  early-formed groups identified from their large magnitude gaps
  (either $\Delta m_{\rm 12} \geq 2.0$ or $\Delta m_{\rm 14} \geq
  2.5$) represent a small fraction (18\% for F14 and 8\% for F12) of
  the overall population of early-formed systems. This is especially
  noticeable in the high-mass regime.

%
%
\item Finally, 
  Fig.~\ref{alifig09} shows that in comparison to conventional fossils
(i.e. F$_{\rm 12}$ groups), 
the  F$_{\rm 14}$
groups
  identified based on $\Delta m_{\rm 14} \geq 2.5$,  
  predominantly correspond to systems with halo masses $M(R_{\rm 200}) \leq
10^{14}\,h^{-1}\,$M$_{\odot}$.

This makes the criterion $\Delta m_{\rm 12} \geq 2.0$ marginally more
efficient than $\Delta m_{\rm 14} \geq 2.5$ in identifying massive
early-formed systems.

\end{enumerate}

These results depend to some extent on the employed semi-analytic
model in our current analysis, and the statistics might change if one
uses different semi-analytical model of galaxy formation. 

Physical prescriptions such as galaxy merging, supernova and AGN
feedback used in such models are somewhat different from one another.
Furthermore, superfluous mergers may result from the algorithm used
for the identification of haloes in the Millennium DM simulation, and
this may affect the merger rates calculated from various studies,
including ours, using these catalogues \citep[e.g.][]{Genel09}.
Though merging is the most important process affecting galaxies in
groups, there are other physical processes such as ram pressure
stripping, interactions and harassment, group tidal field, and gas
loss, that are not fully characterised by current semi-analytic
models. 

 This is partially due to the limited spatial resolution of the
  Millennium simulation.  The new release of the current simulation,
  i.e. Millennium-II Simulation might help to address some of the
  above issues.  The latter has 5 times better spatial resolution and
  125 times better mass resolution \citep{Boylan09}.  Future
  semi-analytic models based upon high resolution simulations,
incorporating such effects would be worth employing in a similar
investigation to find better observational indicators of the ages of
galaxy groups and clusters.

\section*{Acknowledgments}
The Millennium Simulations used in this paper was carried out by the
Virgo Supercomputing Consortium at the Computing Center of the
Max-Planck Society in Garching. The semi-analytic galaxy catalogues
used in this study are publicly available at
http://galaxy-catalogue.dur.ac.uk:8080/MyMillennium/.  The Millennium
Gas Simulations were carried out at the Nottingham HPC facility, as
was much of the analysis required by this work.
The SDSS-DR4 group catalogue of \citet{Yang07} used in this study is publicly available at http://www.astro.umass.edu/~xhyang/Group.html.

AAD gratefully acknowledges Graham Smith, Malcolm Bremer and the
anonymous referee for helpful discussions.  The 2dFGRS group catalogue
data \citep{Yang05,Yang07}
used in this study was kindly provided by Frank C. van den Bosch
and X. Yang.

\label{lastpage}

\end{document}